\pdfoutput=1
\documentclass[10pt,conference]{IEEEtran}
\usepackage{graphicx}
\usepackage[normalem]{ulem}
\usepackage{amsmath}
\usepackage[colorlinks=true,urlcolor=blue, citecolor=blue, linkcolor=blue, plainpages=false]{hyperref}   
\usepackage{amsmath,amssymb,physics,dsfont}
\usepackage{tikz}
\usetikzlibrary{quantikz2}
\usepackage{comment}
\usepackage{mathtools, cuted}
\usepackage{caption}
\usepackage{placeins}
\usepackage{dblfloatfix}

\title{Hardware Realization of a Hamiltonian Simulation Algorithm for Time-Domain Maxwell's Equations}
\author{
\IEEEauthorblockN{
Gautam Sharma,
Apurva Tiwari,
Niladri Gomes,
Jezer Jojo,
J.~Eric Bracken,
Jay Pathak
}
\IEEEauthorblockA{
Synopsys Inc., USA
}
}
\date{}

\begin{document}
\maketitle

\begin{abstract}
We present the first quantum-hardware implementation of a Hamiltonian simulation algorithm that produces signed vector-field solutions to the time-domain Maxwell's equations using a Schr\"odingerisation-based approach. The electromagnetic fields are discretized using finite-difference operators, and the resulting non-unitary matrices are mapped to Bell-basis Trotter blocks, enabling efficient circuit construction. We introduce a measurement procedure that retrieves not only field amplitudes, but also physical directions of the electric and magnetic field values at select spatial points. Implementing this logic on quantum hardware relies on relative-phase–based sign reconstruction. 
Numerical results obtained using IonQ QPU, show good agreement with analytical solutions of benchmark problems in two dimensions; and on simulators, in three dimensions. We further extend our approach to compute fields scattered from simple bodies, by enforcing appropriate boundary conditions.
Our work lays the foundational steps towards realizing quantum‑hardware solutions for computational electromagnetics.
\end{abstract}

\begin{IEEEkeywords}
Hamiltonian Simulation, Schrödingerisation, Maxwell's equation, Electromagnetic vector fields
, Bell-basis decomposition, Trapped ion quantum hardware.
\end{IEEEkeywords}

\section{Introduction}
Quantum computing is emerging as a powerful computational approach for tackling a broad class of problems in computational sciences, seeking to advance classical simulation techniques limited by prohibitively high computational costs that scale unfavorably with system size. Many engineering problems of practical interest are governed by partial differential equations (PDEs), which arise naturally in fluid dynamics, electromagnetics, acoustics, elasticity, plasma physics among others. Significant progress has been made in recent years in the development of quantum algorithms for numerically solving hyperbolic PDEs, with proposed applications to wave equations \cite{costa2019quantum,suau2021practical,sato2024hamiltonian,arute2019quantum, wright2024noisy,lubasch2025quantum} and, in particular, to electromagnetic field simulation
\cite{bui2022alternative,tezuka2025quantum,jin2024quantum,ma2024schr,vahala2020effect,vahala2020unitary,vahala2021one,vahala2021two,nguyen2024solving,colella2025time}. While several quantum PDE solvers have
been introduced, their practical realizations on quantum hardware remain
limited. Existing quantum algorithms for solving PDEs typically fall into two broad categories.
Variational quantum algorithms have been demonstrated for solving PDEs, but
they are constrained by classical optimization bottlenecks, hardware noise
sensitivity, and limited scalability in terms of problem size \cite{cerezo2021variational,lubasch2020variational}. Non-variational approaches, such as Hamiltonian simulation or quantum lattice-based methods, avoid classical optimization bottlenecks but suffer from high circuit depths and  have so far been applied primarily to scalar equations, or to the evaluation of scalar observables derived from more complex systems \cite{gomes2025hamiltonian,tiwari2025algorithmic,sato2024hamiltonian,wawrzyniak2025quantum}. Extending such methods to physically relevant vector-valued systems remains an open challenge.

Electromagnetic field simulation is particularly challenging in this regard. Maxwell’s equations govern coupled vector fields whose components have both magnitude and direction, whereas quantum measurements can recover the output state information only up to a global phase. As a result, even when the underlying time evolution is simulable, extracting physical electromagnetic field values remains nontrivial. Furthermore, the coupling between electric and magnetic fields significantly increases the effective problem size compared to scalar PDEs. Additional complexities arise from non‑unitary evolution operators induced by spatial discretization and from boundary conditions. While several quantum algorithms have been developed to handle non‑unitary dynamics via dilation or linear‑combination techniques \cite{berry2017quantum,PhysRevLett.133.230602,an2023linear}, their practical implementation remains limited by high circuit‑depth and precision requirements.

At the same time, a favorable feature of Maxwell’s equations is their linear structure, which avoids the
difficulties associated with non-linear PDEs \cite{tennie2025quantum}. Moreover, many practical electromagnetic applications do not require knowledge of the field everywhere in the computational domain, but rather at selected spatial locations or regions of interest. This observation
suggests that full field reconstruction or global state tomography is both unnecessary and inefficient for realistic problem sizes. Instead, a viable
quantum electromagnetic solver must provide a means of recovering locally defined vector-field observables with minimal measurement overhead.

In this work, we present a Schrödingerisation-based Hamiltonian simulation
framework for solving Maxwell’s equations in the time domain on quantum computers. The electromagnetic fields are discretized on a staggered Yee grid, converting the curl equations into a system of coupled ordinary differential equations. These equations are then recast into a form compatible
with Schrödingerisation. To enable scalability to larger grids and practical hardware implementations, the resulting evolution operators are implemented using an efficient Bell-basis decomposition, together with a Trotterized time-integration scheme, to avoid the high overhead associated with Pauli-based representations.

Beyond representing the field dynamics, our framework is designed to
support the extraction of physically meaningful electromagnetic observables
on quantum hardware. By focusing on local measurements at selected grid
points, rather than full reconstruction of the field over the entire
domain, the approach remains compatible with large spatial grids.

The main contributions of this work can be summarized as follows:

\begin{itemize}

\item \textbf{Schrödingerisation-based Hamiltonian simulation of vector-valued Maxwell equations:}
We develop a Schrödingerisation-based Hamiltonian simulation algorithm for solving the time-domain Maxwell’s equations in 
two and three dimensions, discretized on structured rectangular
grids and incorporating perfect electric conductor (PEC) and perfect magnetic conductor (PMC) boundary conditions,
and embedded scatterers.

\item \textbf{Quantum measurement and sign‑resolved reconstruction of electromagnetic field components:}
We present a quantum measurement strategy that enables recovery of both the
magnitudes and physical directions (signs) of electromagnetic vector-field
components at specific grid points, allowing physically meaningful field
values to be obtained directly from quantum hardware.

\item \textbf{Hardware realization on  quantum processors:}
We demonstrate the proposed framework through implementations of
two-dimensional Maxwell problem on IonQ quantum
hardware platform. 

\end{itemize}

%
%Design and analysis of modern electromagnetic systems demand highly
%accurate electromagnetic simulations. In this context, time‑domain
%formulations of Maxwell’s equations, such as finite‑difference
%time‑domain (FDTD) methods, are particularly useful due to their
%conceptual simplicity and their ability to capture both transient and
%broadband behavior. However, achieving higher accuracy generally requires increasingly fine spatial discretizations in order to resolve the
%relevant geometric and physical features of the problem.
%For such refined grids, numerical stability conditions impose
%correspondingly smaller time step sizes, which in turn lead to a very
%large number of time steps for long‑time evolutions. As a result, the
%computational cost of highly accurate time‑domain electromagnetic
%simulations rapidly exceeds the memory and runtime capabilities of
%currently available classical computing platforms. These challenges are
%well documented in the numerical analysis literature, particularly in
%multi‑dimensional settings and in problems involving complex geometries
%or large spatial domains
%\cite{chen2012discontinuous,weinan2007heterogeneous,teixeira2023finite}.

% \item{\textbf{Benchmarking with Ansys Lumerical}: We have also benchmarked our quantum algorithms against classical simulation of Maxwell's equations.}

% \item{\textbf{Optimized implementation of Bell-Decomposition}: To Add?}

\section{Problem Setup}\label{sec:problem_setup}
This section describes the formulation of Maxwell’s equations under the physical
assumptions relevant to our simulations. We present the governing equations,
define the computational domain and boundary conditions, and introduce the
finite-difference time-domain (FDTD) discretization used to obtain a
finite-dimensional representation of the electromagnetic field evolution.

\subsection{Governing Equations}
We consider the time-domain Maxwell curl equations in differential form, in a homogeneous, lossless,
and isotropic medium with constant permittivity $\epsilon$ and permeability
$\mu$. In the absence of impressed electric or magnetic sources, these
equations take the following form,
\begin{subequations}
\label{eq:maxwell}
\begin{align}
\frac{\partial \mathbf{E}}{\partial t}
    &= \frac{1}{\epsilon}\,\nabla \times \mathbf{H}, \\
\frac{\partial \mathbf{H}}{\partial t}
    &= -\frac{1}{\mu}\,\nabla \times \mathbf{E}.
\end{align}
\end{subequations}
Here $\mathbf{E}$ denotes the electric field and $\mathbf{H}$ the magnetic
field vector.  Throughout this work, we consider source-free electromagnetic propagation. This
assumption results in a homogeneous first-order linear evolution system, which provides a convenient starting point for the discretization.

\subsection{Discretization Scheme: FDTD}
We model the electromagnetic fields on a three-dimensional rectangular domain
\begin{align}
    \Omega = [0,L_x]\times[0,L_y]\times[0,L_z],
\end{align}
which is discretized using a uniform Cartesian grid in all spatial directions.
The computational domain $\Omega$ is divided into
$N_x \times N_y \times N_z$ cells with grid spacings
$(\Delta x, \Delta y, \Delta z)$. We define the Maxwell’s curl equations over a staggered grid in space using the standard Yee finite-difference time-domain (FDTD) scheme \cite{Yee}. Specifically, the electric-field components are defined at
\begin{align}
(E_x, E_y, E_z) :
(i+\tfrac12,j,k),\;
(i,j+\tfrac12,k),\;
(i,j,k+\tfrac12),
\end{align}
while the magnetic-field components are defined at
\begin{align}
(H_x, H_y, H_z) :&
(i,j+\tfrac12,k+\tfrac12),\;
(i+\tfrac12,j,k+\tfrac12),\; \nonumber \\
&
(i+\tfrac12,j+\tfrac12,k).
\end{align}
This staggered arrangement ensures second-order accuracy in space and preserves
the intrinsic divergence-free structure of Maxwell’s equations. 

Due to the staggered Yee-grid arrangement, field components defined at half-integer locations along a given spatial direction $l \in \{x,y,z\}$ are naturally supported on only $N_l - 1$ grid points along that axis. For later convenience in quantum implementation, we extend such fields to length $N_l$ by padding with a zero value. This padding does not alter the physical evolution, as the added degree of freedom is fixed to zero by construction. For example, the electric-field component $E_x$, which is defined at
locations $(i+\tfrac{1}{2},j,k)$ on the Yee grid, is defined on $N_x - 1$ points along the $x$-direction. We pad this representation with a zero so that all field components can be treated consistently as vectors of length $N_x$.

\paragraph{Component-wise Maxwell equations.}
Maxwell's curl equations Eq.~\eqref{eq:maxwell}, have the standard component form
\begin{align}\label{eq:maxwell_components}
\frac{\partial E_x}{\partial t}
 &= \frac{1}{\epsilon}
    \left( \frac{\partial H_z}{\partial y}
         - \frac{\partial H_y}{\partial z} \right), \quad
\frac{\partial H_x}{\partial t}
 = -\frac{1}{\mu}
    \left( \frac{\partial E_z}{\partial y}
         - \frac{\partial E_y}{\partial z} \right), \nonumber\\[2mm]
\frac{\partial E_y}{\partial t}
 &= \frac{1}{\epsilon}
    \left( \frac{\partial H_x}{\partial z}
         - \frac{\partial H_z}{\partial x} \right), \quad
\frac{\partial H_y}{\partial t}
 = -\frac{1}{\mu}
    \left( \frac{\partial E_x}{\partial z}
         - \frac{\partial E_z}{\partial x} \right), \nonumber\\[2mm]
\frac{\partial E_z}{\partial t}
 &= \frac{1}{\epsilon}
    \left( \frac{\partial H_y}{\partial x}
         - \frac{\partial H_x}{\partial y} \right), \quad
\frac{\partial H_z}{\partial t}
 = -\frac{1}{\mu}
    \left( \frac{\partial E_y}{\partial x}
         - \frac{\partial E_x}{\partial y} \right).
\end{align}

\paragraph{Finite-difference approximation.}
We approximate the spatial derivatives in the curl operators using central finite
differences which is consistent with the staggering of field components on the
Yee grid. Because each electric and magnetic-field component is defined at a
specific offset location, the discrete curl operators naturally couple
neighboring field values symmetrically.

For instance, using central-differences in the $y$- and $z$-directions, the time derivative of $E_x$ at $(i+\tfrac{1}{2},j,k)$ is
\begin{align}
\label{eq:central_diff}
&\frac{\partial E_x(i+\tfrac12,j,k)}{\partial t}  \nonumber \\
&= \frac{1}{\epsilon}
\Big[
H_z(i+\tfrac12,j+\tfrac12,k)
-
H_z(i+\tfrac12,j-\tfrac12,k)
\Big] \nonumber \\
&-
\frac{1}{\epsilon}
\Big[
H_y(i+\tfrac12,j,k+\tfrac12)
-
H_y(i+\tfrac12,j,k-\tfrac12)
\Big].
\end{align}

All remaining spatial derivatives appearing in
Eq.~\eqref{eq:maxwell_components} are discretized in an analogous manner, with
each derivative taken using central differences along the appropriate
coordinate direction. 

\paragraph{Matrix formulation.}
After discretization in space, Maxwell’s equations reduce to a finite-dimensional
first-order system of ordinary differential equations. Let $\vec{E}$ and $\vec{H}$
denote the stacked vectors of all discrete electric- and magnetic-field
components defined on the Yee grid. Each vector has dimension $3N$, where
$N = N_x N_y N_z$ is the number of grid cells in the computational domain.

Collecting all degrees of freedom yields the semi-discrete system
\begin{align}
\label{eq:semi_discrete}
\frac{\partial}{\partial t}
\begin{pmatrix}
\vec{E} \\[1mm]
\vec{H}
\end{pmatrix}
=
\begin{pmatrix}
0 & \mathcal{C}_E \\[1mm]
\mathcal{C}_H & 0
\end{pmatrix}
\begin{pmatrix}
\vec{E} \\[1mm]
\vec{H}
\end{pmatrix},
\end{align}
where $\mathcal{C}_E : \mathbb{R}^{3N} \to \mathbb{R}^{3N}$ and
$\mathcal{C}_H : \mathbb{R}^{3N} \to \mathbb{R}^{3N}$ are sparse matrices encoding
the discrete curl operators acting on the magnetic and electric-field
components, respectively.

At the algebraic level, the curl operator admits the matrix representation
\begin{align}
\label{eq:nabla_EH}
\nabla \times (\cdot) =
\begin{bmatrix}
0  & -\dfrac{\partial}{\partial z} & \dfrac{\partial}{\partial y} \\
\dfrac{\partial}{\partial z} & 0 & -\dfrac{\partial}{\partial x} \\
-\dfrac{\partial}{\partial y} & \dfrac{\partial}{\partial x} & 0
\end{bmatrix}(\cdot),
\end{align}
when acting on vector-valued fields. Although, both
$\mathcal{C}_E$ and $\mathcal{C}_H$ share the same algebraic structure but differ
in their precise matrix entries  in the discrete form due to the use of central finite-difference operators Eq.~\eqref{eq:central_diff} on the staggered grid. This distinction is reflected by the separate subscripts $E$ and $H$.

The resulting system \eqref{eq:semi_discrete} is linear and first-order in time,
with a generator that is generally non-Hermitian. This
finite-dimensional formulation serves as the starting point for the
Schrödingerisation procedure introduced in the next section.

\subsection{Boundary Conditions}

In this work, we consider both perfect
electric conductor (PEC) and perfect magnetic conductor (PMC) boundary
conditions, applied either at the exterior of the computational domain or on
internal embedded interfaces. The choice of these boundaries yields clear physical constraints without introducing additional absorbing-layer complexity.

A perfect electric conductor enforces the vanishing of the tangential electric
field at the boundary,
\begin{align}
\vec{n} \times \mathbf{E} = 0,
\end{align}
where $\vec{n}$ denotes the outward unit normal to the boundary surface. This
condition represents an idealized conducting interface that prevents tangential
electric fields at the boundary.

The complementary boundary condition is the perfect magnetic conductor, which
enforces the vanishing of the tangential magnetic field,
\begin{align}
\vec{n} \times \mathbf{H} = 0.
\end{align}
PMC boundaries may be interpreted as magnetic-wall conditions and arise naturally
in symmetry planes, reduced-dimensional formulations, or idealized truncations.

In the finite-difference formulation, boundary conditions are enforced by
constraining specific field components and by modifying the discrete curl
operators locally near the boundary. For example, at the plane $y=0$, with PMC boundary enforces the vanishing of the tangential magnetic-field components $H_x$ and $H_z$. On the Yee grid, these tangential components are defined at locations adjacent to
the boundary and are removed as active degrees of freedom. This
constraint is enforced by introducing antisymmetric ghost-field extensions
across the boundary. For example, the tangential component $H_z$ may be extended
such that
\[
H_z(i+\tfrac12,-\tfrac12,k) = -\,H_z(i+\tfrac12,\tfrac12,k),
\]
ensuring a zero average magnetic field at the boundary. As a result, finite-difference stencils involving $\partial H_z/\partial y$ are locally modified
near $y=0$, yielding truncated or one-sided expressions while preserving the
required boundary condition. This approach applies uniformly to boundaries on the domain boundaries as well
as to internal scatterers embedded within the
computational region. 

Before proceeding to the quantum implementation, we note that Schrödingerisation‑based quantum simulations of Maxwell’s equations discretized on a Yee grid have also been explored in Refs.~\cite{jin2024quantum,ma2024schr}. Algorithmically our implementation differs from these approaches primarily in the implementation of the Bell-basis decomposition. While the Bell‑basis decompositions employed in Ref.~\cite{jin2024quantum} implements specific operator blocks associated with spatial derivatives, our implementation forms a global construction applied uniformly across the full Hamiltonian, rather than being tailored to individual curl‑derivative terms as we will show in the next section.

\section{Quantum Theory and Methods}\label{sec:quantum_theory}
Building on the semi-discretized Maxwell system Eq.~\eqref{eq:semi_discrete}, we now develop a quantum representation suitable for simulation on a gate-based quantum computer. The discrete evolution operator associated with the FDTD update is generally non-unitary due to the boundary treatments. To accommodate this, we adopt the Schrödingerisation framework \cite{PhysRevLett.133.230602}, which provides a Hamiltonian-simulation–compatible representation of the non-unitary update without enlarging the state dimension. 

\subsection{Schrödingerisation}

We begin with the semi-discretized Maxwell system obtained in
Eq.~\eqref{eq:semi_discrete}, written in the generator form
\begin{align}\label{eq:maxwell_generator_form}
    \frac{d u(t)}{dt} = A\,u(t),
\end{align}
where
\[
u =
\begin{pmatrix}
\vec{E}\\ \vec{H}
\end{pmatrix},
\qquad
A =
\begin{pmatrix}
0 & \mathcal{C}_E \\
\mathcal{C}_H & 0
\end{pmatrix}.
\]

While $A$ is real and linear, it does not in general generate a unitary time
evolution. The following steps convert this non-unitary dynamics into a
form suitable to Hamiltonian simulation. Although motivated here by Maxwell’s
equations, the procedure applies to any linear system arising from the spatial
discretization of a linear PDE.

We decompose the generator as
\begin{align}
A = H_1 + iH_2,
\quad
H_1 = \tfrac12(A + A^\dagger),
\quad
H_2 = \tfrac1{2i}(A - A^\dagger),
\end{align}
with $H_1$ and $H_2$ Hermitian. Introducing an auxiliary real variable $p$, we
define the lifted state
\begin{align}
v(t,p) = e^{-p} u(t),
\end{align}
with initial data extended smoothly for $p<0$. The lifted variable satisfies
\begin{align}
\frac{\partial v}{\partial t}
= -H_1\,\partial_p v + iH_2 v.
\end{align}

Taking the Fourier transform in $p$ decouples the dynamics into a family of
Schrödinger equations,
\begin{align}\label{eq:schro_implementation}
\frac{d}{dt}\,\tilde v(t,\xi)
=
i\bigl(\xi H_1 + H_2\bigr)\tilde v(t,\xi),
\end{align}
where each generator $\xi H_1 + H_2$ is Hermitian. The original solution $u(t)$ is
recovered from the lifted system by evaluating $v(t,p)$ at sufficiently large
positive $p$, yielding
\begin{align}
u(t) = e^{p}\,v(t,p),
\end{align}
with $p$ chosen above a spectral bound determined by $H_1$. Details of the
recovery procedure follow standard Schrödingerisation arguments can be found
in \cite{PhysRevLett.133.230602}.

\subsection{Bell-Basis Decomposition and Circuit Implementation}
Now we will present the Bell-basis decomposition followed by its circuit implementation.

\textbf{Dimension requirements and padding:}
We ensure that the grid dimensions of the computational domain satisfy  
\[
\{N_x, N_y, N_z\} = \{2^{n_x}, 2^{n_y}, 2^{n_z}\},
\]
so that the generator $A$ has dimension $6N \times 6N$ with $N = N_x N_y N_z$.  Therefore, we pad the matrices with zeros to embed the full generator $A$ into an $8N \times 8N$ matrix, which is compatible with the Bell-basis tensor structure.  Next, we decompose the generator into hermitian matrices $H_1$ and $H_2$.

\textbf{Choice of basis:} For trotterization we must decompose the matrices $H_1$ and $H_2$ into tensor-product blocks.  
A Pauli decomposition is possible but not scalable for large discretizations, so we  use the Bell-basis decomposition \cite{sato2024hamiltonian, hu2024quantum} instead, which yields significantly lower circuit depth for derivative-operator matrices.

\textbf{Tensor-product decomposition:}
We express each tensor-product term in the decompositions of $H_1$ and $H_2$ using the
rank‑one matrices
\[
\sigma_{00}=\ketbra{0}{0},\;
\sigma_{01}=\ketbra{0}{1},\;
\sigma_{10}=\ketbra{1}{0},\;
\sigma_{11}=\ketbra{1}{1},
\]
together with $I_2$.  
A generic tensor-product term takes the form
\[
S = e^{i\lambda}\bigotimes_{k=1}^{n} \sigma_{a_kb_k},
\]
where $n = \log_2(N)$, $\lambda$ comes from the coefficients and $(a_k,b_k)\in\{0,1\}^2$ specifies which matrix acts on the
$k$‑th qubit.  
By construction, every such $S$ appears together with its adjoint $S^\dagger$ both
in $H_1$ and $H_2$.

\textbf{Reduction to Bell basis:}
To isolate the nontrivial matrices, we first ignore the factors
$\sigma_{00}$, $\sigma_{11}$, and $I_2$, and consider only the positions where
$\sigma_{01}$ or $\sigma_{10}$ appear.  
Let these positions define a subset of $m$ qubits out of $n$.  
If at position $k$ we have:
\begin{align*}
&\sigma_{01}=\ketbra{0}{1} \;\Rightarrow\; (a_k,b_k)=(1,0), \nonumber \\
&\sigma_{10}=\ketbra{1}{0} \;\Rightarrow\; (a_k,b_k)=(0,1),
\end{align*}
then the tensor product over this subset reduces to the rank‑one operator
\[
T = \ketbra{a}{b},\qquad 
|a\rangle = |a_1\ldots a_n\rangle,\quad
|b\rangle = |b_1\ldots b_n\rangle.
\]
Since $T^\dagger=\ketbra{b}{a}$ also appears, the Hermitian combination is
\[
S' = T + T^\dagger
    = \ketbra{a}{b} + \ketbra{b}{a}.
\]
Every such operator can be decomposed as 
\[
S' = 
\frac{\ket{a}+\ket{b}}{\sqrt{2}}
\frac{\bra{a}+\bra{b}}{\sqrt{2}}
-
\frac{\ket{a}-\ket{b}}{\sqrt{2}}
\frac{\bra{a}-\bra{b}}{\sqrt{2}}.
\]

\textbf{Bell-basis embedding:}
As shown in Lemma~1 of \cite{hu2024quantum}, there exists an operator $O$ acting on these $m$ qubits such that
\[
O\ket{0}\ket{1}^{m-1}=\tfrac{\ket{a}+\ket{b}}{\sqrt{2}},\qquad
O\ket{1}^m=\tfrac{\ket{a}-\ket{b}}{\sqrt{2}}.
\]
This yields the Bell-basis decomposition
\[
S' = O\!\left(Z\otimes\ketbra{1}{1}^{\otimes (m-1)}\right)O^\dagger.
\]

\textbf{Extension back to $S$:}
To recover the full tensor term $S$, we simply re‑insert the
$\sigma_{00}$, $\sigma_{11}$, and $I_2$ factors on the qubits where they
originally appeared:
\[
S = O\!\left(
Z\otimes \ketbra{1}{1}^{\otimes(m-1)}
\;\bigotimes_{k=1}^{n-m}\sigma_{a_k a_k}
\right)O^\dagger.
\]
These extra factors act as like controls and do not affect the
bell‑basis structure.

\textbf{Circuit implementation of $e^{iSt}$:}
Using the Bell-basis decomposition,
\begin{align}
    e^{i S t}\label{eq:circ_expiST}
= O\left(\mathrm{CRZ}_{n}\right)O^\dagger. 
\end{align}

For the full operator $S$, the controlled-$R_Z$ gate acquires
\emph{additional controls} which are determined as:
\begin{itemize}
    \item qubits with $\sigma_{01}$ or $\sigma_{10}$  
        → standard Bell-basis controls,
    \item qubits with $\sigma_{00}$  
        → control conditioned on state $|0\rangle$,
    \item qubits with $\sigma_{11}$  
        → control conditioned on state $|1\rangle$,
    \item qubits with $I_2$  
        → no control.
\end{itemize}
Thus the same qubits that carried the original $\sigma_{a_kb_k}$
matrices become the controls for the $e^{i S t}$ circuit, with control
values determined directly by the type of $\sigma_{ab}$ at that position.

\textbf{Schrödingerisation circuit:}
After constructing circuits for the individual exponentiated tensor‑product blocks $e^{iSt}$, we assemble approximations to the full evolutions $e^{iH_1 t}$ and $e^{iH_2 t}$ using Trotter product formulas. These unitary blocks are then incorporated into the Schrödingerisation framework described in Eq.~\eqref{eq:schro_implementation}, which lifts the original non‑unitary dynamics to a unitary evolution suitable for quantum simulation. The resulting Schrödingerisation circuit follows directly from this construction, and its implementation proceeds analogously to that used in the quantum advection‑equation solver of Ref.~\cite{hu2024quantum}.

\begin{figure*}[!tb]
    \centering
    \includegraphics[width=\textwidth]{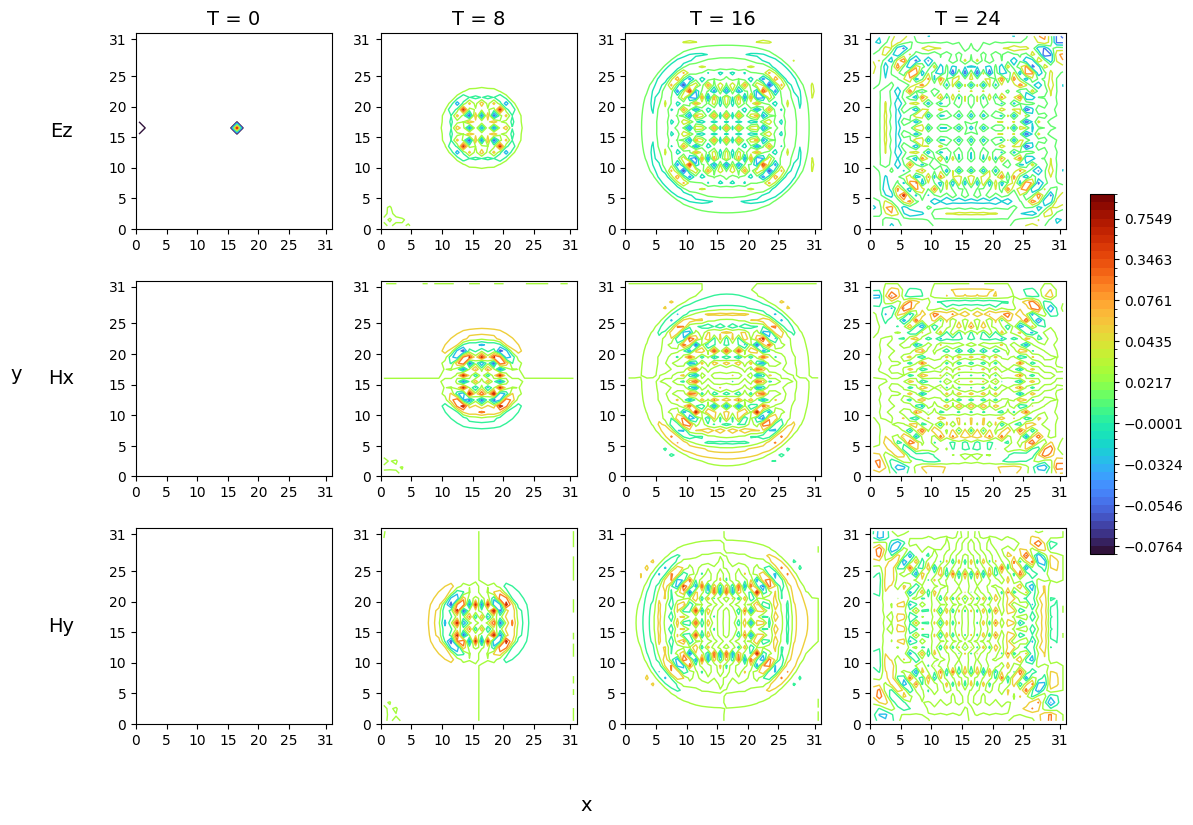}
    \caption{Simulation results for the 2D Maxwell’s equations on an empty $32 \times 32$ grid with PMC boundary conditions on all edges and PEC boundaries on top and bottom surfaces. From top to bottom, the rows show the evolution of the field components $E_z$, $H_x$, and $H_y$, respectively. Columns correspond to snapshots at times $T = 0$, $8$, $16$, and $24$.}
    \label{fig:2D_sim_nobody}
\end{figure*}

\subsection{Measuring Negative field Values}\label{subsection:negative_values}
Quantum computers can recover the final statevector only up to a global phase, which is a fundamental limitation of quantum measurement. In our setting, where the solution consists of vector fields, this limitation means that we can access only the \emph{relative} signs of the field components produced by the quantum algorithm, but not their \emph{absolute} physical signs. To overcome this, we apply the following procedure.

We apply a constant offset to one selected component of the initial electric
field,
\begin{align}
    E_k \;\mapsto\; E_k + C,
\end{align}

where the constant \(C\) is chosen sufficiently large so that the shifted
component remains strictly non‑negative throughout the entire evolution.
The value of \(C\) is determined \emph{a priori} based on bounds implied by
the initial field configuration and the evolution conditions. 

For example, if the simulation is initialized with a localized impulse of
magnitude \(E_{\max}\) in the \(E_z\) component at a single grid point, while
all other field components are initialized to zero and no external source
terms are present, then energy conservation implies that the most negative
value attainable by \(E_z\) during the evolution is \(-E_{\max}\). In this
case, choosing any constant \(C > E_{\max}\) guarantees that the shifted
field \(E_z + C\) remains strictly positive for all times.

This offset does not affect the remaining electric‑ or magnetic‑field
components and serves solely to fix the sign of the selected field component
during the quantum simulation. Upon completion of the simulation, the
original physical field is recovered by removing the offset,
\begin{align}
    E_k \;\mapsto\; E_k - C,
\end{align}
thereby restoring the correct magnitude and sign of the electric‑field  component.
Once the absolute sign (i.e., direction) of this reference component is recovered, the absolute signs of all remaining electric and magnetic field components can be determined consistently using their relative phases relative to this field. 

\subsection{Gate Complexity}

For a computational grid of size $N = N_x N_y N_z$, the Bell‑basis embedding and padding described in the previous section lead to a total Hilbert space dimension of $8N$. Accordingly, the number of system qubits required to represent the discretized electromagnetic fields is
\[
n = \log_2(8N) = \log_2 N + 3.
\]
In addition, a constant number of ancilla qubits is used to implement the Schrödingerisation procedure; the precise number of these ancillae, denoted by $n_a$, depends on the desired precision.

Each exponentiated tensor‑product block $e^{iSt}$ appearing in the trotterized approximations of $e^{iH_1t}$ and $e^{iH_2t}$ consists of the Bell‑basis transformations $O$ and $O^{\dagger}$ together with a multi‑controlled $R_Z$ gate Eq.~\eqref{eq:circ_expiST}. The operators $O$ and $O^{\dagger}$ contribute a circuit depth that scales linearly with the number of system qubits. The dominant non‑Clifford component is the multi‑controlled $R_Z$ gate, whose implementation requires a number of two‑qubit (CNOT) gates that is upper‑bounded by $\mathcal{O}(n^2)$, while the associated single‑qubit gate count scales as $\mathcal{O}(n)$.

After feeding these trotterized blocks into the Schrödingerisation circuit, the resulting circuit has a two‑qubit gate count that is upper bounded by $\mathcal{O}(n^2 N_a)$, where $N_a = 2^{n_a}$ \cite{hu2024quantum}, Lemma 5.

Throughout this paper we are following first-order trotterization. The fact that the exponentiation of different parts of the generator matrix are implemented separately, leads to increase in the trotteriation error of the implementation. The accuracy as well as resulting circuit depth depends on the number of tensor‑product blocks appearing in this decomposition, with fewer blocks leading directly to more efficient implementations. 

\section{Simulation Results}
We present two‑ and three‑dimensional numerical results obtained from classical execution of the proposed quantum circuits with Qiskit, serving to validate the algorithm prior to hardware implementation.

\subsection{Maxwell's Equation in 2D}

\subsubsection{Without Internal PEC Scatterer (PMC Boundary Conditions)}\label{subsubsection:no_body}

In the two--dimensional setting, we model an electromagnetic cavity that is uniform along the $z$--direction and sufficiently thin compared to its lateral dimensions. Under the 2D TE$_z$ formulation, we choose to retain only the field components $(E_z, H_x, H_y)$ by assuming invariance along the $z$--axis. The cavity is terminated by PEC boundaries on the upper and lower $z$--faces, while the outer $(x,y)$ edges impose PMC boundary conditions.

\begin{figure*}[!tb]
    \centering
    \includegraphics[width=\textwidth, keepaspectratio]{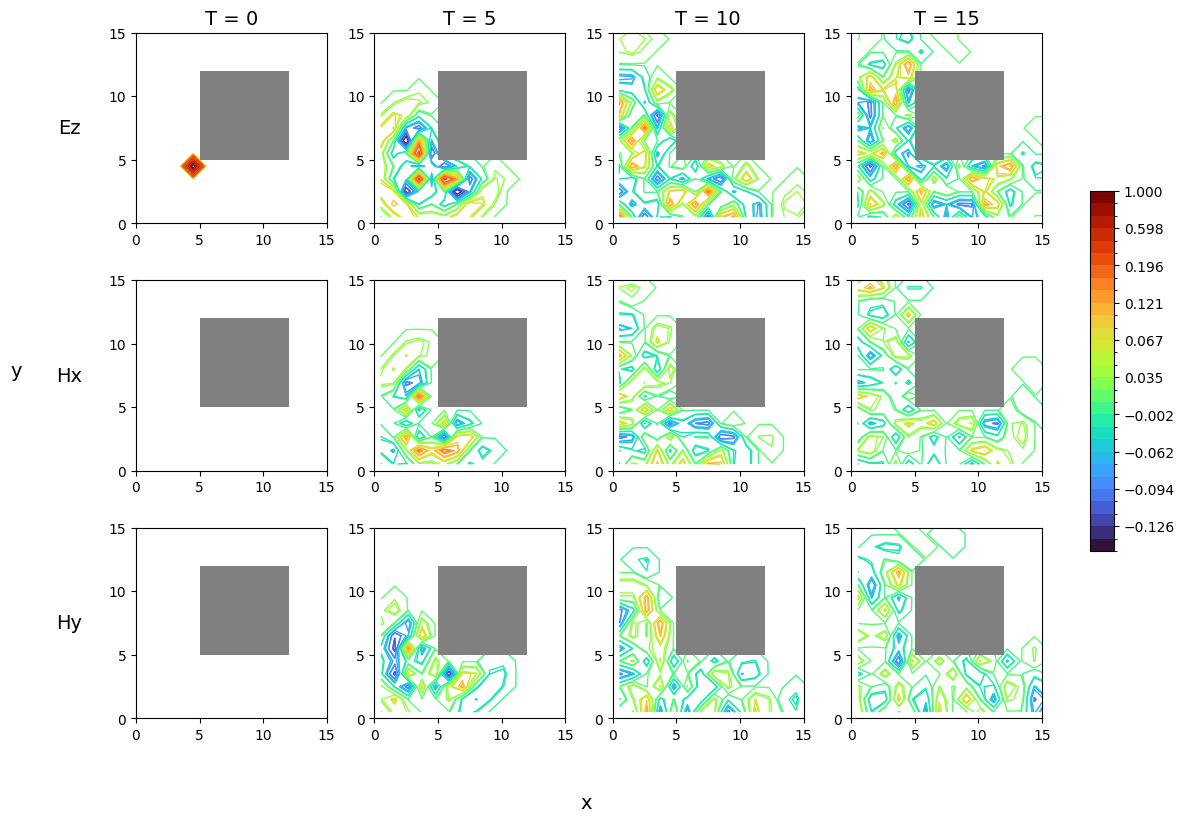}    
    \caption{Simulation results for the 2D Maxwell’s equations on a
$16 \times 16$ grid with an internal scatterer and PMC boundary
conditions on all outer edges.
From top to bottom, the rows show the evolution of the field components
$E_z$, $H_x$, and $H_y$, respectively.
Columns correspond to snapshots at times $T = 0$, $5$, $10$, and $15$.}
\label{fig:2D_sim_body}
\end{figure*}

Following Sec. \ref{sec:problem_setup} under these assumptions, Maxwell's equations reduce to the coupled system
\begin{align}\label{eq:2D_Maxwell}
\frac{\partial E_z}{\partial t}
    =  \left(
         \frac{\partial H_y}{\partial x}
         - \frac{\partial H_x}{\partial y}
       \right), 
\frac{\partial H_x}{\partial t}
    = -\frac{\partial E_z}{\partial y},
\frac{\partial H_y}{\partial t}
    = \frac{\partial E_z}{\partial x},
\end{align}
where we have chosen $\epsilon = \mu = 1$. Also, the computational domain is discretized on a 2D uniform Cartesian grid of size $N_x \times N_y$ with node locations $(i, j)$. The fields are defined on the staggered 2D grid as 
\begin{align}
(E_z, Hx, Hy) :
(i,j),\; (i,j+\tfrac12),\; (i+\tfrac12,j).
\end{align}

This staggering yields central finite--difference approximations for all spatial derivatives. Since the 2D setup of the problem doesn't involve all the derivatives in Eq.~\ref{eq:maxwell_generator_form}, we get rid of the unncessary terms along with the source term, so that the generator matrix for 2D Maxwell is 
\begin{align}
    A_{2D} =
\begin{pmatrix}
0 & \nabla_{E_1} & \nabla_{E_2} & 0 \\
\nabla_{H_1} & 0 & 0 & 0 \\
\nabla_{H_2} & 0 & 0 & 0 \\
0 & 0 & 0 & 0
\end{pmatrix}, 
\end{align}
where the derivative terms have the following form
\begin{align}\label{eq:A2_components}
    &\nabla_{E_1} = -\frac{d}{dy}\otimes I^n,  \quad
    \nabla_{E_2} = I^n\otimes \frac{d}{dx}, \nonumber \\
    &\nabla_{H_1} = -\frac{d}{dy}\otimes I^n, \quad
    \nabla_{H_2} = I^n\otimes \frac{d}{dx},
\end{align}
where the discretized form of the derivatives is implemented with central differences Eq.~\eqref{eq:central_diff}. Notice that by padding with additional zero valued rows and columns, in this form the $A_{2D}$ matrix has dimensions $4N_xN_y \times 4N_xN_y$, hence suitable for Bell basis decomposition given $N_x$ and $N_y$ are some powers of 2. Once the form of generator is established we decompose it in the Bell basis and implement the evolution using Schrödingerisation as described in Sec \ref{sec:quantum_theory}. 

We present the simulation results in Fig.~\ref{fig:2D_sim_nobody}. The simulation is carried out on a $N_x \times N_y = 32 \times 32$ grid, corresponding to $\log_2(4N_xN_y) = 12$ system and one ancilla qubit for Schrodingerisation. As the initial condition, we excite  the electric field Ez with an impulse of unit-amplitude at $(x_0,y_0) = (16,16)$  while the magnetic fields are initially set to zero everywhere. 
We evolve the system up to $T = 24$ using a trotter time step of $\mathrm{d}t = 0.01$, resulting in a total of 2400 time steps. Over this evolution, the initial delta‑like electric field spreads symmetrically and reflects off the domain boundaries. The fields $H_x$ and $H_y$ exhibit similar behavior, propagating outward and reflecting from the edges. The progression of this wave is observed at times $T = 0, 8, 16,$ and $24$. 

We evaluate the $\ell_2$--norm error between the quantum trotterized evolution and the analytical solution using explicit form of $\exp(At)$ for $(E_z, H_x, H_y)$. Errors for two trotter time steps, $\mathrm{d}t = 0.1$ and $\mathrm{d}t = 0.01$, are listed in the table \ref{tab:trotter_error}. The error follows the expected first‑order Trotter scaling $\mathcal{O}(t\, \mathrm{d}t)$, decreasing with smaller time steps, growing linearly in time, and exhibiting consistent behavior across all field components.

\begin{table}[h!]
\centering
\begin{tabular}{|c|ccc|ccc|}
\hline
 & \multicolumn{3}{c|}{$\mathrm{d}t = 0.01$}
 & \multicolumn{3}{c|}{$\mathrm{d}t = 0.1$} \\ \hline
Time & Ez & Hx & Hy & Ez & Hx & Hy \\ \hline
8  & 0.015  & 0.011 & 0.011 & 0.15  & 0.113 & 0.113 \\ \hline
16 & 0.067  & 0.024 & 0.024 & 0.29  & 0.209 & 0.212 \\ \hline
24 & 0.1394 & 0.078 & 0.078 & 0.443 & 0.325 & 0.325 \\ \hline
\end{tabular}
\caption{Trotterization error at different times $T$ for $\mathrm{d}t = \{0.01,0.1\}$.}
\label{tab:trotter_error}
\end{table}

\subsubsection{With Internal PEC Scatterer (PMC Boundary Conditions)}

We next examine wave propagation in the presence of an internal scatterer. A square perfectly electric conducting body $\mathcal{B}$ is embedded inside the computational domain. As in the empty-cavity case, the outer domain boundaries impose PMC boundary conditions, while the scatterer enforces PEC conditions on its top and bottom $z$-faces and PMC conditions on its lateral edges.

In the empty region of the 2D grid, the same governing equations in
Eq.~\eqref{eq:2D_Maxwell} remain valid. The field components are defined on the same staggered Yee grid as before, except that degrees of freedom corresponding to points lying inside $\mathcal{B}$ are excluded. The boundary conditions introduced by the scatterer are incorporated through local modifications of the discrete derivative operators $\nabla_{E_1}$, $\nabla_{E_2}$, $\nabla_{H_1}$, and $\nabla_{H_2}$. Despite these modifications, the resulting generator matrix $A_{2D}$ retains dimension $4N_xN_y \times 4N_xN_y$.

The simulation results in the presence of the scatterer are shown in
Fig.~\ref{fig:2D_sim_body}. The chosen grid size $N_x \times N_y = 16 \times 16$ requires $\log_2(4N_xN_y) = 10$ system qubits and one ancilla qubit. Moreover, the scatterer body has dimensions $(N_x/2,N_y/2)$ with its center aligned with the center of the grid.
To avoid excitation within the scatterer, the initial electric-field impulse with unit-amplitude 
$E_z$ is placed in the empty region at $(x_0, y_0) = (N_x/4, N_y/4)$, while the magnetic-field components $H_x$ and $H_y$ are initially set to zero everywhere, as in the empty-grid case. 

The system is evolved with time step $dt = 0.1$ up to a final time $T = 15$, corresponding to 150 Trotter steps. The resulting field evolution exhibits reflections from both the outer domain boundaries and the internal scatterer. This behavior is evident in the field snapshots at $T = 0, 5, 10,$ and $15$. Consistent with the empty-domain simulations, the $\ell_2$-norm error remains
of order $10^{-1}$ and grows approximately linearly with the evolution time $T$, as expected for first-order Trotterization.

\begin{figure}
    \centering
    \includegraphics[width=\columnwidth , keepaspectratio]{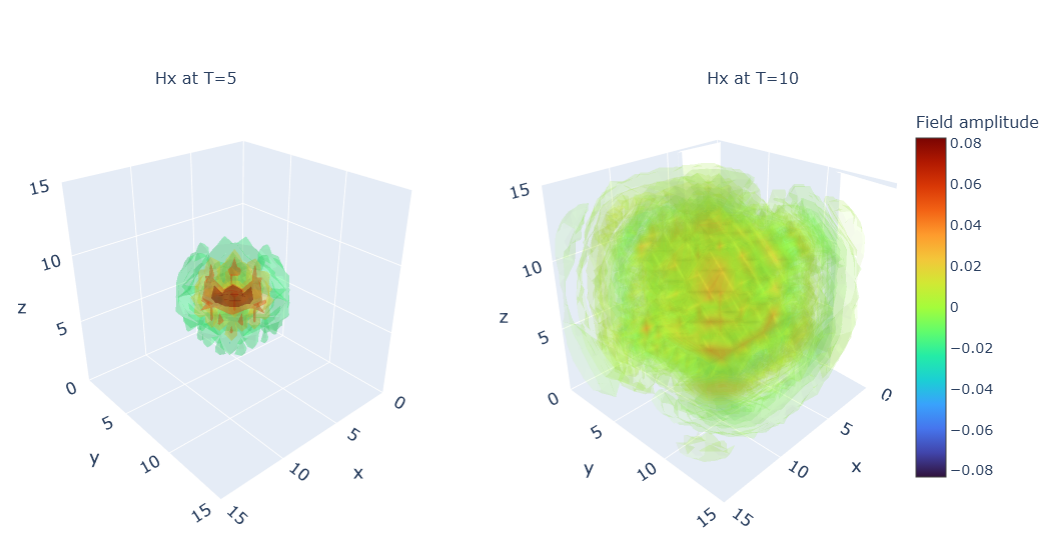}    
    \caption{Three-dimensional evolution of the magnetic-field component $H_x$
shown as an isosurface on a $16 \times 16 \times 16$ grid with PMC boundary
conditions. The fields are excited by an initially localized point source in the $E_z$
component at the center of the domain, $(x_0,y_0,z_0)=(8,8,8)$.}
    \label{fig:3D_Hx}
\end{figure}

\begin{figure*}[tb]
    \centering
    \includegraphics[width=\textwidth, keepaspectratio]{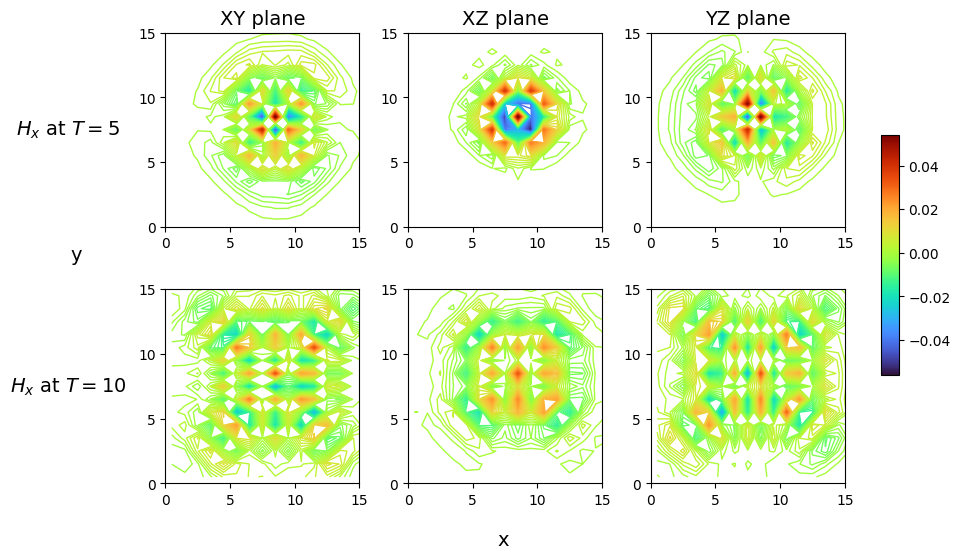}    
    \caption{Orthogonal two-dimensional cross sections of the $H_x$ field taken
through the center of the domain in the $xy$, $xz$, and $yz$ planes at  times $T=5,10$.}
    \label{fig:3D_cross_sections}
\end{figure*}

\subsection{Maxwell's Equation in 3D}
Finally, we present simulation results for Maxwell's equations on a
three-dimensional empty grid. Perfect magnetic conductor (PMC) boundary conditions are imposed on all surfaces of the computational domain. Unlike the two-dimensional case, the full three-dimensional formulation requires retaining all curl-derivative terms appearing in
Eq.~\eqref{eq:semi_discrete}, hence the overall problem setup follows the description in Sec.~\ref{sec:problem_setup}.

The 3D simulation results shown in
Figs.~\ref{fig:3D_Hx} and \ref{fig:3D_cross_sections}
illustrate the evolution of the $H_x$ field, including an isosurface view and the corresponding two-dimensional cross sections taken along the three orthogonal planes passing through the domain center.
The simulations are performed on a $16 \times 16 \times 16$ grid, which corresponds to $\log_2(8N_xN_yN_z)=15$ system qubits and one ancilla qubit. The initial condition consists of a point excitation of unit-amplitude in the $E_z$ field at the center of the domain, $(x_0,y_0,z_0) = (8,8,8)$, while all remaining field components are initialized to zero.

The system is evolved using a time step $dt = 0.1$ for 100 time steps,
corresponding to a total evolution time $T = 10$. The resulting two-dimensional cross section in the $xy$-plane closely resembles the $H_x$ field evolution observed in the 2D empty-grid case shown in Fig.~\ref{fig:2D_sim_nobody}, providing a qualitative consistency check between the 2D and 3D implementations. As in the previous simulations, the $\ell_2$-norm error remains of order
$10^{-1}$ for the chosen $dt$.

% \subsection{Benchmarking with Lumerical}

\section{Hardware Implementation}
We demonstrate our algorithm on the IonQ Forte Enterprise system, a
36‑qubit trapped‑ion quantum computer. The circuits produced by our Bell‑basis Trotterization are far too deep to be executed directly, so they must be transpiled and compressed
before submission. To address this, we use Adaptive Approximate Quantum Compiling (ADAPT‑AQC)
\cite{jaderberg2026variational} to reduce circuit depth while maintaining high fidelity with the ideal evolution.

\subsubsection{Circuit Compression with ADAPT-AQC}
Adaptive Approximate Quantum Compiling (ADAPT-AQC) \cite{jaderberg2026variational}
is an adaptive variational algorithm for compressing quantum circuits.  
ADAPT-AQC adds the approximate circuit by adding a two-qubit 
unitary at a time, instead of using a fixed ansatz. At each iteration a location is selected that reduces the compilation cost the most. The target matrix product state(MPS) representation---is kept fixed, and the ansatz is constructed such that the composite circuit $V^\dagger(\vec{\theta})U$ approaches identity.

The algorithm proceeds iteratively in following steps:
\begin{enumerate}
    \item Convert the target circuit to a Matrix product state (MPS) using a tensor-network simulator.
    \item At iteration $n$, consider all allowed two-qubit placements for the next unitary.
    \item For each candidate position, evaluate the cost gradient $\|\nabla C\|$ 
          using the Loschmidt Echo cost function $C = 1 - |\langle 0 | V^\dagger(\vec{\theta}) U |0\rangle|^2|$,
          computed via MPS contractions.
    \item Insert the two-qubit unitary at the position giving the largest gradient (steepest cost reduction).
    \item Optimise its parameters with \texttt{rotoselect}, then sweep through all ansatz
          parameters with \texttt{rotosolve}.  
\end{enumerate}

This process repeats until the fidelity threshold is reached or no further improvement is obtained.  Since the algorithm does not assume a specific ansatz pattern, it is able to discover shallow, hardware-efficient structures that are not obvious from the original circuit. In our workflow, the entire uncompressed circuit is provided directly to
ADAPT-AQC as the compilation target. This produces a significantly shallower circuit that approximates the full
target unitary to high fidelity.

Finally, we describe the specific ADAPT-AQC configuration used in our circuit-compression workflow. Following the general ADAPT-AQC framework introduced in Ref.~\cite{jaderberg2026variational}—in which layers are added adaptively using incremental structural learning (ISL) and parameters are updated with rotosolve/rotoselect—we employ the ``general gradient'' strategy described in the paper for selecting new two-qubit blocks. Concretely, we set the compiler to use the ISL method with gradient-based pair selection, a rotoselect tolerance of $10^{-4}$, a rotosolve tolerance of $10^{-3}$, and a rotosolve update frequency of 10. The maximum number of two-qubit gates was restricted to 100, and at most 10 layers were modified during each iteration. We also enabled the reuse priority heuristic (\texttt{reuse\_priority\_mode}=\texttt{``qubit''}) with a memory window of 30 steps. These settings match the gradient-driven expansion strategy described in Appendix~A of Ref.~\cite{jaderberg2026variational}, where the selection criterion is determined by the magnitude of the cost-function gradient with respect to candidate two-qubit unitaries.

All ADAPT-AQC compilation steps were executed using Qiskit’s MPS simulator backend, employing an MPS truncation threshold of $10^{-16}$ and a maximum bond dimension of 1000. The simulator was wrapped using the \texttt{mps\_sim\_with\_args()} utility, enabling efficient MPS updates in line with the caching approach described in Appendix~B of Ref.~\cite{jaderberg2026variational}. Each ADAPT-AQC run was initialised with a single-layer product-state ansatz (\texttt{initial\_single\_qubit\_layer=True}), consistent with the recommended low-depth initialisation strategy outlined in the paper. The compiled circuit is obtained as \texttt{result.circuit} from the \texttt{AdaptCompiler} and used as the compressed representation of the original circuit.

\begin{figure*}[tb]
    \centering
    \includegraphics[width=\textwidth, keepaspectratio]{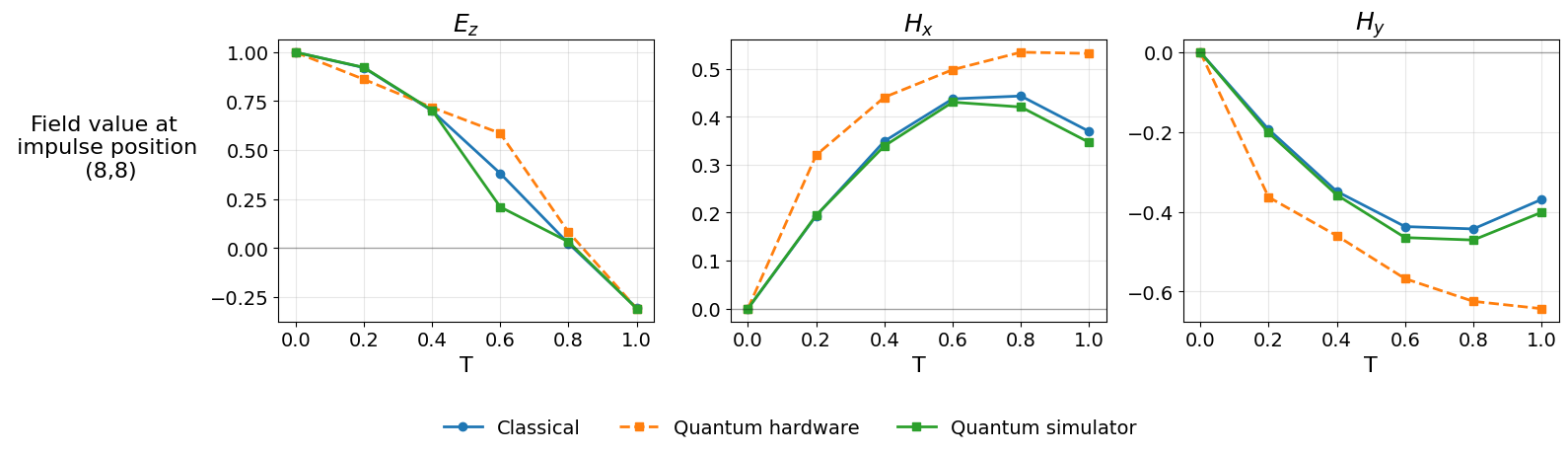}    
    \caption{Comparison of field values measured at the center of the
$16 \times 16$ computational grid, $(8,8)$, where the initial electric-field
excitation is applied.
Time evolution of the field components $E_z$, $H_x$, and $H_y$ is shown for
classical simulation, noise-free quantum simulation, and quantum hardware
execution.
The noise-free quantum results closely follow the analytical solution,
while the hardware results qualitatively reproduce the correct temporal
behavior despite the presence of device noise.}
    \label{fig:2D_hardware}
\end{figure*}

\subsubsection{Transpilation and Error Mitigation}

Once the circuit depth was reduced to a level suitable for execution, we transpiled each circuit for the IonQ backend using Qiskit’s built-in transpilation pipeline. In our workflow, we simply specified the IonQ backend, allowing Qiskit to handle the conversion to the device-supported gate set automatically, without manually selecting basis gates or using IonQ-specific native-gate workflows \cite{ionq_qiskit_sdk_docs}. 

For error mitigation, we relied entirely on IonQ’s default settings. IonQ backends automatically apply debiasing---a compiler-level technique that generates symmetric circuit variants and aggregates their results---for all jobs executed with at least 500 shots \cite{ionq_qiskit_sdk_docs}. Since our experiments were run with $2^{13}$ shots, which is well above this threshold, debiasing was applied automatically, and we did not override or configure any error-mitigation parameters.

\subsection{Results}

We present hardware results obtained using IonQ’s trapped-ion quantum
device in Fig.~\ref{fig:2D_hardware}.
The physical problem corresponds to a two-dimensional empty computational
grid with PMC boundary conditions applied on the domain edges and PEC
boundary conditions on the top and bottom faces, as described in
Sec.~\ref{subsubsection:no_body}.

Unlike classical simulation of the quantum algorithm, execution on quantum
hardware yields the final quantum state only up to a global phase.
To recover the physically meaningful signs (field directions), we employ
the sign-reconstruction procedure described in
Sec.~\ref{subsection:negative_values} by choosing the offset to the electric field $E_z$ to be $C=1$.
Importantly, our objective is not to reconstruct the full electromagnetic
field over the entire grid, but rather to measure the values of selected
field components at specific spatial locations of interest.

The signed value of a field component at a given grid point is obtained using a small number of expectation--value measurements. We first measure the magnitude $M_{\mathrm{ref}}$ of a reference field (chosen as $E_z$), whose absolute sign is fixed using the sign--reconstruction procedure, followed by the magnitude $M$ of the target field. The relative sign between the two amplitudes is then extracted via an interference--based Hadamard measurement \cite{nielsen2010quantum}, by comparing expectation values proportional to $(M_{\mathrm{ref}}+M)^2$ and $(M_{\mathrm{ref}}-M)^2$. If $(M_{\mathrm{ref}}+M)^2$ exceeds $(M_{\mathrm{ref}}-M)^2$, the fields have the same sign; otherwise, their signs are opposite. This approach incurs constant measurement overhead per target field and grid point and scales efficiently without requiring full--field reconstruction.

The grid size for the hardware implementation is $16 \times 16$, which
corresponds to $10$ system qubits and one ancilla qubit. For this problem instance, naive Trotterized circuits result in depths on
the order of $\sim 4\times10^4$. By applying ADAPT-AQC, we reduce the
overall circuit depth to approximately $\sim 200$ across the different
circuits used in the measurement procedure, while keeping the number of
CNOT gates below $150$ for all circuits.

The initial condition consists of a localized non-zero electric-field
excitation of unit amplitude in the $E_z$ component at $(x_0,y_0) = (8,8)$, while the
magnetic-field components $H_x$ and $H_y$ are initialized to zero
everywhere.
The evolution is carried out with a time step $dt = 0.1$ and executed for
up to $100$ time steps, corresponding to a total propagation time of
$T = 1$.

We focus our analysis on the field values measured at the center of the
computational grid, where the initial electric-field excitation is applied.
As shown in Fig.~\ref{fig:2D_hardware}, noise-free quantum simulations closely
match the analytical solution, while hardware results, though affected by
device noise, remain in good qualitative agreement. The remaining discrepancies
are primarily attributable to hardware noise and finite sampling. To the best of our knowledge, this constitutes the first experimental
demonstration of a quantum algorithm that computes and measures a
time-evolving electromagnetic vector field on quantum hardware.

\section{Conclusion}

We have proposed and implemented a quantum algorithm for time‑dependent electromagnetic field evolution governed by Maxwell’s equations. The approach relies on a Schrödingerisation‑based Hamiltonian simulation technique for enabling implementation of the inherently non-unitary dynamics arising from the discretized governing equations.  By combining it with a Bell-basis decomposition along with first order trotterization, we obtain a structured and hardware-compatible framework for time-domain electromagnetic evolution. Crucially, we prescribe a measurement procedure that allows the recovery of signed electric and magnetic field components at selected spatial locations, avoiding the need for full state reconstruction, and enabling physically meaningful observables to be obtained directly from the quantum hardware.

We validate our proposed framework through implementations on both on  noise‑free classical backends, and on a trapped‑ion quantum device, with significant circuit‑depth reductions achieved via ADAPT‑AQC. In the noise-free setting, the simulated dynamics  closely reproduce the corresponding analytical solutions. While the hardware results are inevitably subject to device noise and finite sampling,  they qualitatively reproduce the correct field evolution nonetheless, and demonstrate the feasibility of the proposed algorithm with NISQ devices.

The current study focuses on a restricted set of electromagnetic configurations, including empty computational domains with ideal PEC and PMC boundary conditions, simple rectangular scatterers, and moderate spatial resolutions. Extending the algorithm to more complex
electromagnetic problems, to include explicit source terms, dielectric bodies,
more realistic scattering geometries, and larger spatial grids—will substantially
increase the problem complexity and is a natural direction for future work.

From a quantum‑algorithmic perspective, there is scope for improvement in the construction of resource‑efficient circuits. In particular, higher‑order
Trotterization schemes, alternative Hamiltonian‑simulation techniques such as
qubitization, and more structured operator factorizations to enable longer
time evolutions, better error control and larger problem sizes while being implementable on contemporary quantum hardware.

\bibliographystyle{unsrt}
\bibliography{references}

\begin{thebibliography}{10}

\bibitem{costa2019quantum}
Pedro~CS Costa, Stephen Jordan, and Aaron Ostrander.
\newblock Quantum algorithm for simulating the wave equation.
\newblock {\em Physical Review A}, 99(1):012323, 2019.

\bibitem{suau2021practical}
Adrien Suau, Gabriel Staffelbach, and Henri Calandra.
\newblock Practical quantum computing: Solving the wave equation using a
  quantum approach.
\newblock {\em ACM Transactions on Quantum Computing}, 2(1):1--35, 2021.

\bibitem{sato2024hamiltonian}
Yuki Sato, Ruho Kondo, Ikko Hamamura, Tamiya Onodera, and Naoki Yamamoto.
\newblock Hamiltonian simulation for hyperbolic partial differential equations
  by scalable quantum circuits.
\newblock {\em Physical Review Research}, 6(3):033246, 2024.

\bibitem{arute2019quantum}
Frank Arute, Kunal Arya, Ryan Babbush, Dave Bacon, Joseph~C Bardin, Rami
  Barends, Rupak Biswas, Sergio Boixo, Fernando~GSL Brandao, David~A Buell,
  et~al.
\newblock Quantum supremacy using a programmable superconducting processor.
\newblock {\em nature}, 574(7779):505--510, 2019.

\bibitem{wright2024noisy}
Lewis Wright, Conor Mc~Keever, Jeremy~T First, Rory Johnston, Jeremy Tillay,
  Skylar Chaney, Matthias Rosenkranz, and Michael Lubasch.
\newblock Noisy intermediate-scale quantum simulation of the one-dimensional
  wave equation.
\newblock {\em Physical Review Research}, 6(4):043169, 2024.

\bibitem{lubasch2025quantum}
Michael Lubasch, Yuta Kikuchi, Lewis Wright, and Conor Mc~Keever.
\newblock Quantum circuits for partial differential equations in fourier space.
\newblock {\em Physical Review Research}, 7(4):043326, 2025.

\bibitem{bui2022alternative}
Nicolas Bui, Alain Reineix, and Christophe Guiffaut.
\newblock Alternative quantum circuit implementation for 2d electromagnetic
  wave simulation with quasi-pec modeling.
\newblock In {\em 2022 IEEE MTT-S International Conference on Electromagnetic
  and Multiphysics Modeling and Optimization}, 2022.

\bibitem{tezuka2025quantum}
Hiroyuki Tezuka and Yuki Sato.
\newblock Quantum algorithm for electromagnetic field analysis.
\newblock {\em arXiv preprint arXiv:2510.03596}, 2025.

\bibitem{jin2024quantum}
Shi Jin, Nana Liu, and Chuwen Ma.
\newblock Quantum simulation of maxwell’s equations via
  schr{\"o}dingerisation.
\newblock {\em ESAIM: Mathematical Modelling and Numerical Analysis},
  58(5):1853--1879, 2024.

\bibitem{ma2024schr}
Chuwen Ma, Shi Jin, Nana Liu, Kezhen Wang, and Lei Zhang.
\newblock Schr$\backslash$" odingerization based quantum circuits for maxwell's
  equation with time-dependent source terms.
\newblock {\em arXiv preprint arXiv:2411.10999}, 2024.

\bibitem{vahala2020effect}
George Vahala, Linda Vahala, Min Soe, and Abhay~K Ram.
\newblock The effect of the pauli spin matrices on the quantum lattice
  algorithm for maxwell equations in inhomogeneous media.
\newblock {\em arXiv preprint arXiv:2010.12264}, 2020.

\bibitem{vahala2020unitary}
George Vahala, Linda Vahala, Min Soe, and Abhay~K Ram.
\newblock Unitary quantum lattice simulations for maxwell equations in vacuum
  and in dielectric media.
\newblock {\em Journal of Plasma Physics}, 86(5):905860518, 2020.

\bibitem{vahala2021one}
George Vahala, Linda Vahala, Min Soe, and Abhay~K Ram.
\newblock One-and two-dimensional quantum lattice algorithms for maxwell
  equations in inhomogeneous scalar dielectric media i: theory.
\newblock {\em Radiation Effects and Defects in Solids}, 176(1-2):49--63, 2021.

\bibitem{vahala2021two}
George Vahala, Min Soe, Linda Vahala, and Abhay~K Ram.
\newblock One-and two-dimensional quantum lattice algorithms for maxwell
  equations in inhomogeneous scalar dielectric media. ii: Simulations.
\newblock {\em Radiation Effects and Defects in Solids}, 176(1-2):64--72, 2021.

\bibitem{nguyen2024solving}
Nam Nguyen and Richard Thompson.
\newblock Solving maxwells equations using variational quantum imaginary time
  evolution.
\newblock {\em arXiv preprint arXiv:2402.14156}, 2024.

\bibitem{colella2025time}
E~Colella, F~Moglie, V~Mariani Primiani, and G~Gradoni.
\newblock Time domain field simulations on quantum computers via
  riemann-silberstein formulation.
\newblock In {\em 2025 International Applied Computational Electromagnetics
  Society Symposium (ACES)}, pages 1--2. IEEE, 2025.

\bibitem{cerezo2021variational}
Marco Cerezo, Andrew Arrasmith, Ryan Babbush, Simon~C Benjamin, Suguru Endo,
  Keisuke Fujii, Jarrod~R McClean, Kosuke Mitarai, Xiao Yuan, Lukasz Cincio,
  et~al.
\newblock Variational quantum algorithms.
\newblock {\em Nature Reviews Physics}, 3(9):625--644, 2021.

\bibitem{lubasch2020variational}
Michael Lubasch, Jaewoo Joo, Pierre Moinier, Martin Kiffner, and Dieter Jaksch.
\newblock Variational quantum algorithms for nonlinear problems.
\newblock {\em Physical Review A}, 101(1):010301, 2020.

\bibitem{gomes2025hamiltonian}
Niladri Gomes, Gautam Sharma, and Jay Pathak.
\newblock Hamiltonian simulation for solving the advection equation with
  arbitrary velocity field.
\newblock In {\em 2025 IEEE International Conference on Quantum Computing and
  Engineering (QCE)}, volume~1, pages 1849--1858. IEEE, 2025.

\bibitem{tiwari2025algorithmic}
Apurva Tiwari, Jason Iaconis, Jezer Jojo, Sayonee Ray, Martin Roetteler, Chris
  Hill, and Jay Pathak.
\newblock Algorithmic advances towards a realizable quantum lattice boltzmann
  method.
\newblock In {\em 2025 IEEE International Conference on Quantum Computing and
  Engineering (QCE)}, volume~1, pages 2441--2451. IEEE, 2025.

\bibitem{wawrzyniak2025quantum}
David Wawrzyniak, Josef Winter, Steffen Schmidt, Thomas Indinger, Christian~F
  Jan{\ss}en, Uwe Schramm, and Nikolaus~A Adams.
\newblock A quantum algorithm for the lattice-boltzmann method
  advection-diffusion equation.
\newblock {\em Computer Physics Communications}, 306:109373, 2025.

\bibitem{berry2017quantum}
Dominic~W Berry, Andrew~M Childs, Aaron Ostrander, and Guoming Wang.
\newblock Quantum algorithm for linear differential equations with
  exponentially improved dependence on precision.
\newblock {\em Communications in Mathematical Physics}, 356(3):1057--1081,
  2017.

\bibitem{PhysRevLett.133.230602}
Shi Jin, Nana Liu, and Yue Yu.
\newblock Quantum simulation of partial differential equations via
  schr\"odingerization.
\newblock {\em Phys. Rev. Lett.}, 133:230602, Dec 2024.

\bibitem{an2023linear}
Dong An, Jin-Peng Liu, and Lin Lin.
\newblock Linear combination of hamiltonian simulation for nonunitary dynamics
  with optimal state preparation cost.
\newblock {\em Physical Review Letters}, 131(15):150603, 2023.

\bibitem{tennie2025quantum}
Felix Tennie, Sylvain Laizet, Seth Lloyd, and Luca Magri.
\newblock Quantum computing for nonlinear differential equations and
  turbulence.
\newblock {\em Nature Reviews Physics}, 7(4):220--230, 2025.

\bibitem{Yee}
Kane Yee.
\newblock Numerical solution of initial boundary value problems involving
  maxwell's equations in isotropic media.
\newblock {\em IEEE Transactions on Antennas and Propagation}, 14(3):302--307,
  1966.

\bibitem{hu2024quantum}
Junpeng Hu, Shi Jin, Nana Liu, and Lei Zhang.
\newblock Quantum circuits for partial differential equations via
  schr{\"o}dingerisation.
\newblock {\em Quantum}, 8:1563, 2024.

\bibitem{jaderberg2026variational}
Ben Jaderberg, George Pennington, Kate~V Marshall, Lewis~W Anderson, Abhishek
  Agarwal, Lachlan~P Lindoy, Ivan Rungger, Stefano Mensa, and Jason Crain.
\newblock Variational preparation of normal matrix product states on quantum
  computers.
\newblock {\em Physical Review Research}, 8(1):013081, 2026.

\bibitem{ionq_qiskit_sdk_docs}
IonQ.
\newblock Qiskit sdk documentation.
\newblock \url{https://docs.ionq.com/sdks/qiskit}.
\newblock Accessed: 2026-03-11.

\bibitem{nielsen2010quantum}
Michael~A Nielsen and Isaac~L Chuang.
\newblock {\em Quantum computation and quantum information}.
\newblock Cambridge university press, 2010.

\end{thebibliography}

\end{document}